\documentclass[twocolumn]{aa}
\usepackage{graphicx}
\usepackage{txfonts}
\begin{document}
   \title{Astrochemical confirmation of the 
          rapid evolution of
          massive YSOs 
	  and explanation for the inferred ages of hot cores}


   \author{S. D. Doty,
          \inst{1}
	   E. F. van Dishoeck,
	  \inst{2}
          \and
          J. C. Tan
	  \inst{3}
          }

   \offprints{S. D. Doty, \email{doty@denison.edu}}

   \institute{Department of Physics and Astronomy, Denison University,
              Granville, OH, 43023, USA 
	\and
	     Sterrewacht Leiden, PO Box 9513, 2300 RA Leiden, 
	     The Netherlands
         \and
	     Department of Astronomy, University of Florida, 
	     P. O. Box 112055, Gainesville, FL, 32611, USA
             }

   \date{Received ; accepted }

   \abstract
   {}
   {
   To understand the roles of infall and protostellar evolution on the
   envelopes of massive young stellar objects (YSOs).
   }
   {
   The chemical evolution of gas and dust is traced, including
   infall and realistic source evolution.
   The temperatures are determined self-consistently.
   Both ad/desorption of ices using recent
   laboratory temperature-programmed-desorption measurements are included.
   }
   {
   The observed water abundance jump near $100$ K is reproduced by 
   an evaporation front which moves outward as the luminosity increases.
   Ion-molecule reactions produce water below $100$ K.  
   The age of the source is constrained to $t \sim
   8 \pm 4 \times 10^{4}$ yrs since YSO formation.
   It is shown that the chemical age-dating of hot cores at 
   $\sim \mathrm{few} \times 10^{3} - 10^{4}$ yr and the
   disappearance of hot cores on a timescale of 
   $\sim 10^{5}$ yr is a natural consequence of infall in
   a dynamic envelope and protostellar evolution.  
   Dynamical structures of  $\sim 350 \mathrm{AU}$ 
   such as disks should contain 
   most of the complex second generation species.
   The assumed order of desorption kinetics does not affect these
   results.
   }
   {}


  \titlerunning{
     Effects of Infall and Source Evolution
     }

   \maketitle
%

\section{Introduction}

   The physical and chemical structure of
   massive young stellar objects (YSOs) has been an area of
   increasing study (e.g. 
   Beuther et al.  \cite{Beutheretal2006}) 
   While a wealth of new data and models 
   have become available, many basic open questions remain.  
   For example, the processes and timescales involved in the
   formation and evolution of the high-mass protostellar objects
   and ensuing ``hot core'' phase are not on firm theoretical ground.
   
   Recently, McKee \& Tan (\cite{McKeeTan2003}) introduced the
   first realistic model for the evolution of massive YSOs,
   including a time-dependent accretion rate.
   By combining this with a model for the evolution of the 
   protostar onto the ZAMS, they predicted the time-dependent
   total (source + accretion) luminosity for the lifetime of the
   massive YSO.  

   A test of the evolution is expected to be
   desorption of ices from grain mantles, 
   as the evolving source heats the surrounding envelope.
   Laboratory data using temperature 
   programmed desorption (TPD) have been recently reported
   (see e.g., Fraser et al. \cite{Fraseretal2001}).
   These data allow for
   a detailed analysis of the desorption of
   realistic ice mantles.

   Models have included some, but not all, of these effects.
   While TPD data were included by Viti et al. (\cite{Vitietal2004}) and
   Lintott et al. (\cite{Lintottetal2005}), 
   these were single-point models and/or used 
   approximate source evolution.  
   Source evolution
   was considered by Rodgers \& Charnley (\cite{RodgersCharnley2003})
   and Lee et al. (\cite{LeeBerginEvans2004}),  
   but using low-mass YSO models.
   It is common to assume a static envelope, 
   constant luminosity,
   and standard first-order desorption 
   (e.g., Caselli et al. 
   \cite{CaselliHasegawaHerbst1993}; 
   Nomura \& Millar \cite{NomuraMillar2004}).
Viti \& Williams (\cite{VitiWilliams1999})
    and others 
    have used multi-point models. 
  
   In this letter we describe the first
   model for the chemical evolution in a collapsing massive YSO, including
   the evolution of the central source, infall, and ad/desorption
   of ices from the grains.
   The 
   models are used as a clock for studying the evolution of the
   central source.  
   Here we concentrate
   on the distribution of H$_{2}$O,  since it is
   not only a dominant ice component and important thermal regulator, 
   but also a target of the 
   upcoming Herschel 
   Space 
   Observatory.  We apply the model to the 
   AFGL 2591 envelope as it is one of the best observed sources for chemistry
   in general, and water in particular.


\section{Source and Model}

AFGL 2591 has been 
observed in water
lines over a range of excitation conditions with both 
SWAS 
(Snell et al. \cite{Snelletal2000}) 
and 
ISO 
(Boonman \& van Dishoeck \cite{BoonmanvanDishoeck2003}).  
Boonman et al. (\cite{Boonmanetal2003}) used radiative transfer models to 
constrain the water 
distribution, with
three conclusions:
(1) water ice evaporates at 
$r \sim 10^{16-16.5}$ cm [$T \sim 100$ K in
the static model];
(2) $x(\mathrm{H}_{2}\mathrm{O})_{T>100} \equiv 
n(\mathrm{H}_{2}\mathrm{O})/ n(\mathrm{H}_{2})_{T>100} = 
2 \times 10^{-4}$; and
(3) $x(\mathrm{H}_{2}\mathrm{O})_{T<100} \leq 10^{-8}$.
This 
distribution has been confirmed by recent
H$_{2}^{18}$O observations by van der Tak et al. (\cite{vanderTaketal2006}).

%
   \begin{figure}
   \centering
   \resizebox{\hsize}{!}{\includegraphics{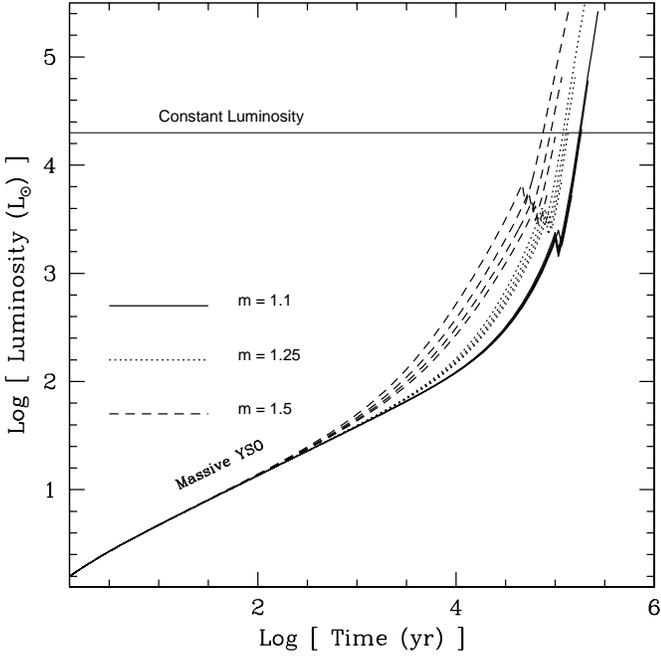}}
      \caption{Evolution of $L(t)$. 
	       The curves within each line type
	       correspond to 
               final stellar masses of 
               $40, 20, 12$, and 8 M$_{\odot}$ 
	       (top to bottom). 
              }
         \label{fig1}
   \end{figure}
%

Van der Tak et al. 
(\cite{vdt2000})
constructed spherically-symmetric models of AFGL 2591 
and found a best fit density power law  
$n(r) \propto r^{-m}$ where $m \sim 1$.
Since the source evolution models of McKee \& Tan (\cite{McKeeTan2003})
break down for $m=1$ exactly as the pressure gradient vanishes
in this limit, we adopt $m=1.1$.  
We take a final
stellar mass of $m_{*,f}=20$ M$_{\odot}$.
All other inputs are taken from 
McKee \& Tan (\cite{McKeeTan2003}) for 
cores.
Consistent with $m \sim 1$, we adopt a logotropic collapse velocity field 
(McLaughlin \& Pudritz \cite{McLaughlinPudritz1997}).  
While chemical and radiative transfer modeling suggest $m=1$, we also
consider an $m=1.5$ polytrope. 

\begin{figure}
   \centering
   \resizebox{\hsize}{!}{\includegraphics{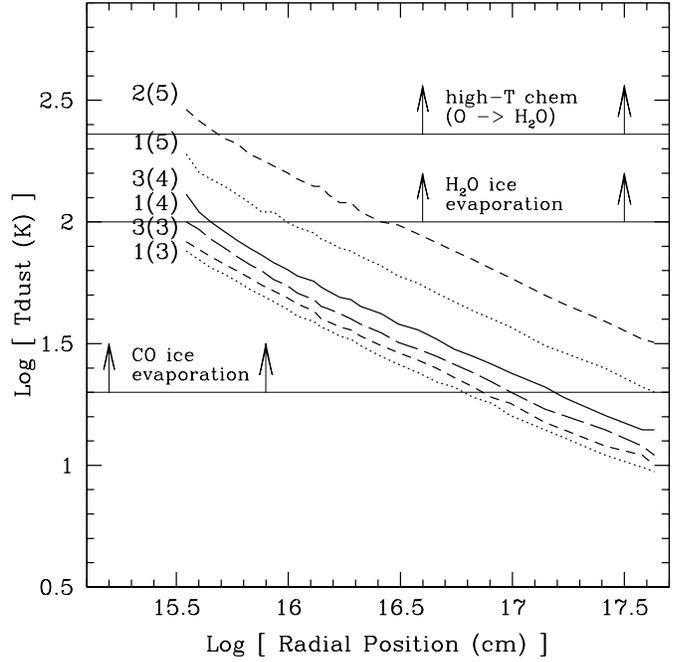}}
      \caption{Temperature distribution as a function of position
               and time for $m=1.1$.  The curves are labeled by
	       the time in yr.
              }
         \label{fig2}
   \end{figure}

Figure \ref{fig1} shows the luminosity evolution, $L(t)$.
The results are almost independant of $m_{*,f}$.
Given $n(r,t)$ and $L(t)$
we solve for 
$T(r,t)$.
Figure \ref{fig2} shows $T_{\mathrm{dust}}(r,t)$.

The chemical model is based upon the best-fit
model of Doty et al. (\cite{Dotyetal2002}) for AFGL 2591, using the same
gas-phase chemical network, cosmic-ray
ionization rate, and initial abundances 
in the cold cloud exterior.  
In particular, since $L \sim 0$ at $t=0$,
the input mantle composition places all H$_{2}$O and CO
on the grain surfaces intially.  
The CO is quickly liberated --
by $3 \times 10^{3}$ yr in the H$_{2}$O evaporation zone, and
by $3 \times 10^{4}$ yr throughout the envelope.

We have made two important
modifications.  First, we allow the material to infall, solving the
chemical evolution in a Lagrangian frame, while simultaneously 
interpolating on $T(r,t)$.  Second we include 
ad/desorption.
For water, we incorporate both
the zeroth-order kinetics of Fraser et al. 
(\cite{Fraseretal2001}), and the first-order 
kinetics of Sandford \& Allamandola (\cite{SandfordAllamandola1988}), 
in turn.
For other species, we follow the usual first-order kinetics.

%
%


\section{Results}

\begin{figure}
   \centering
   \resizebox{\hsize}{!}{\includegraphics{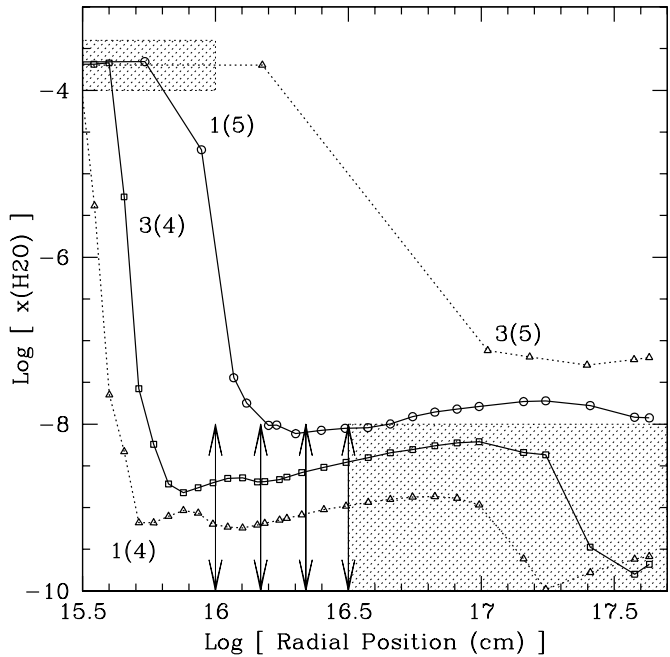}}
      \caption{Water abundance as a function of position
               in AFGL 2591.  The different line types correspond
	       to different times since the formation of the YSO.
	       The shaded areas are the observational constraints.
              }
         \label{fig3}
   \end{figure}

The water abundance 
after including 
infall and source evolution 
for $m=1.1$ is shown in Fig. \ref{fig3}.
The 
abundance
constraints are noted by the shaded regions, while the
region of the observed evaporation step is noted by 
the double arrows.

These results can be understood given the fact that
for $T_{\mathrm{dust}} > 100$ K, water ice evaporates quickly.
As a result, the $T_{\mathrm{dust}}>100$ K region
lies behind an outward propagating evaporation front
at which the water ice evaporates nearly instantaneously from the grain surface.
The evaporation front stops at $r \sim 3 \times 10^{16}$ cm and 
$t \sim 2 \times 10^{5}$ yr, as $m_{*} \rightarrow m_{*,f}$.
For the $m=1.5$ polytrope, this occurs at $t \sim 9 \times 10^{4}$ yr.

The water abundance for $T_{\mathrm{dust}}<100$ K can be
understood in terms of ion-molecule chemistry.
Here CO is dissociated into O.  Proton-transfer
processes this into H$_{3}$O$^{+}$, which then dissociatively
recombines into H$_{2}$O.  The net result is that CO is gradually
transformed into water over time in the cool exterior.  

This combination of effects brackets the
observational constraints shown in Fig. \ref{fig3}.
For $t \leq 3 \times 10^{4}$ yr, the luminosity has not increased to 
where the evaporation front can match the observed size of the region
of enhanced gas-phase water of $r \sim 10^{16-16.5}$ cm.  
For $t > 10^{5}$ yr, the abundance in
the outer envelope is too high primarily due to 
ion-molecule production of water for $T < 100$ K.  
Together this means that a time $t = 3 - 10 \times 10^{4}$ yr 
since YSO formation is
sufficient to liberate water, without warming the entire envelope
or producing significant ion-molecule chemistry.  Results for
an $m=1.5$ polytrope are nearly identical.

%
%

Lower stellar masses yield commensurately longer
times to reach a given desorption radius, 
due to their longer Kelvin-Helmholz timescales.
Fig. \ref{fig1} shows that the observational differences
are small:  $L(t)$ varies
by $< 0.5$ dex for $8 < m_{*,f}(M_{\odot}) < 40$.  
Since desorption occurs around  
$r_{100\mathrm{K}} \sim 4 \times 10^{14}$ cm $(L/3L_{\odot})^{1/2}$,
a stellar mass 5 times smaller leads to a desorption radius
at any time that is only marginally ($<1.8 \times$) smaller. 

\begin{figure}
   \centering
   \resizebox{\hsize}{!}{\includegraphics{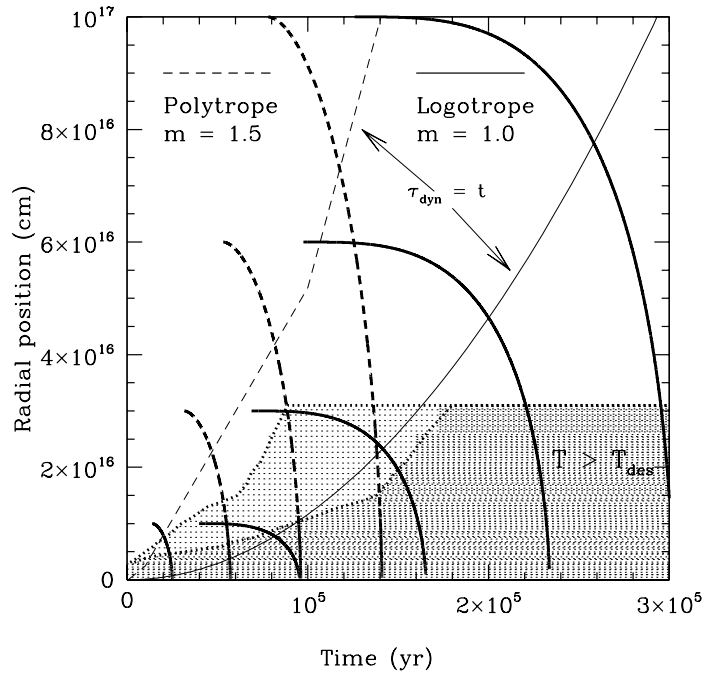}}
      \caption{Infall trajectories of parcels (bold),
	       crossed by lines noting 
	       where dynamics become important (thin).
	       The shaded regions
	       signify desorption (dark = logotrope,
	       light = polytrope). 
              }
         \label{fig4}
   \end{figure}

To understand other implications, in Fig. 
\ref{fig4} we plot the infall trajectories of parcels.
Grain mantles quickly evaporate ($T > T_{\mathrm{des}}$)
in the shaded region 
The lines crossing the trajectories signify where dynamics becomes
important ($\tau_{\mathrm{dyn}} \sim t$).  

We first note that for parcels originating at
$r \leq 10^{16}$ cm, the effects of source evolution
dominate collapse, since $T$ reaches $T_{\mathrm{des}}$
before dynamics become important.  
On the other hand, parcels originating at $r > 3-6 \times 10^{16}$ cm
are dynamic in the mantle evaporation region, due to the fast 
evolution of the central source.
In all cases the gas spends little time above $T_{\mathrm{des}}$.  

\begin{figure}
   \centering
   \resizebox{\hsize}{!}{\includegraphics{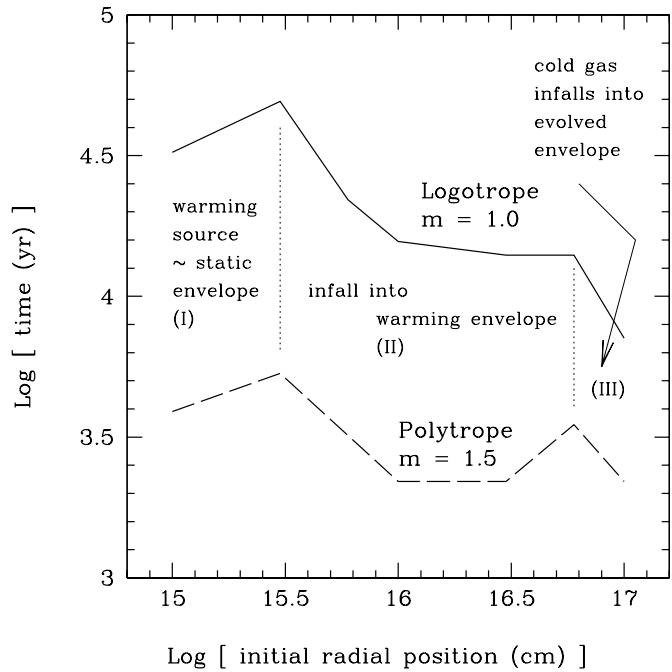}}
      \caption{Time spent above $T_{\mathrm{des}}$ in the extended
               envelope.  
              }
         \label{fig5}
   \end{figure}

To see this more clearly, in Fig. \ref{fig5} we plot the 
time spent by parcels above $T_{\mathrm{des}}$ as a function of their
initial position.
Qualitatively, the time spent in the warm gas
can be understood in three regimes.  (I) At early times and small distances,
the envelope is $\sim$ static and evolution is
dominated by the source.  Parcels that begin further
out spend more time at high temperatures due to their longer infall distance.
(II) An  
intermediate regime exists where both dynamics and source
evolution occur.  
The shorter times at high
temperatures due to the faster infall speeds are somewhat balanced by
the outward progression of the evaporation front.
Finally,
(III) cold parcels from far out
warm as they fall into the completely evolved inner envelope.
In this case, parcels that start progressively further out hit the 
now stationary evaporation front at progressively higher speeds, 
leading to less time at high temperatures.

%

Quantitatively, Fig. \ref{fig5} shows that parcels spend 
$\sim \mathrm{few} \times 10^{3} - 10^{4}$ yr as ``warm'' gas,
independent of their starting point.  
Parcels then enter a region of $< 200$ AU ($< 0.2''$ at 1 kpc).
Such regions are severely beam-diluted and thus nearly invisible.
As a result, chemical clocks
based upon warm gas-phase (``hot-core'') species 
-- especially 
observed with single-dish instruments -- should show
ages of $\sim \mathrm{few} \times 10^{3} - 10^{4}$ yr.
These ages do not sample 
the actual age of the source, but instead
the residence time in the extended warm gas.
The combination of 
infall, evaporation, and
source evolution naturally explains the observed consistency of 
chemically inferred ages of hot cores between sources.

Since the outer envelope remains ``cold'' for $ \sim 10^{5}$ yr, 
we predict
a discrepancy between the ages inferred 
from hot core molecules and from cold gas-phase species.  This is 
consistent with the chemically-inferred ages of
$\sim 4 - 40 \times 10^{3}$ yr for hot cores, and $> 10^{5}$ yr for their
cold surrounding envelopes (e.g. Hatchell et al. \cite{Hatchelletal1998},
Millar et al. \cite{Millaretal1997}).

Taken together, we propose the following picture of high-mass
envelope evolution:  for $t < 10^{4}$ yr, the majority of the
observable envelope is essentially static, and a cold gas-phase chemistry
is seen.  As the YSO evolves and warms the envelope,
a warm gas-phase ``hot-core'' chemistry is produced.
For $t<1-8 \times 10^{4}$ yr (depending upon polytropic index), 
a static hot core of size 
$r \sim 10^{16}$ cm is produced.  After this time, 
the ``hot'' material is dynamic, while much of the cold material
remains $\sim$ static.
The ice-evaporated hot core 
chemistry has a characteristic inferred age of 
$\sim \mathrm{few} \times 10^{3} - 10^{4}$ yr, independent of the
actual age of the YSO.  
Finally, once the source
reaches the main-sequence, it is hot and luminous
enough to ionize the envelope, leading to the disappearance of the
hot core after $t \sim 10^{5}$ yr.

In the scenario above, the infalling material accretes onto a
central structure smaller than a few hundred AU.  This
may be an accretion disk that is diluted in large
single-dish 
observations. 
Indeed, the disk scale 
is given by
$r_d=350 \:{\rm AU} \: 
(\beta/0.007)
(m_{*}/ {\rm 20\: M_\odot})^{1/2}$,
where
$\beta$ is the ratio of rotational to gravitational energy.
For high-mass star-formation, $\beta \sim 0.007$
(Pirogov et al. \cite{Pirogovetal2003}).
This size is consistent with a $\sim$ few hundred AU disk inferred
by van der Tak et al. (\cite{vanderTaketal2006}) toward AFGL 2591.  
The column density of this region should be
$> 10^{22}$ cm$^{-2}$,
suggesting
densities $> 10^{7}$ cm$^{-3}$ and temperatures above 100 K.
With lifetimes of $10^{5}-10^{6}$ yr 
(Cesaroni et al. \cite{Cesaronietal2006}), 
this is the regime where warm, complex, 
second generation chemistry can take place.  As a result, 
future observatories such as eSMA, ALMA, and the EVLA, 
which can probe scales of
$< 0.2''$ will be exquisite tools for uncovering and 
age-dating the complex circumstellar 
chemistry.

Lastly, the first-order desorption kinetics of Sandford \& Allamandola 
(\cite{SandfordAllamandola1988}) yield similar
results to the zeroth order kinetics from TPD data. 
The abundance distributions have the same shape,
but shift inward by $\sim 0.1$ dex in position due to
their 
lower desorption energies.
While the order of desorption kinetics is not
significant here, it may be important elsewhere.


%
%

\section{Conclusions}

We have constructed the first model for the chemical evolution
in a collapsing massive YSO, including the realistic evolution
of the central source as well as the ad/desorption of ices from
grain mantles as the grains infall from the cool exterior into the
warming interior.  This approach allows for a  more realistic  
evaporation of ices from the mantles, as opposed to the 
more parametric assumptions of previous models.  We find that this
approach naturally reproduces and explains the parametric
step-function water distribution inferred observationally.

The water abundance jump is naturally explained as the 
propagation of a thermal desorption front outward from the
YSO as it evolves.  The abundance in the cool outer 
envelope is explained by ion-molecule reactions which 
convert CO into H$_{2}$O.  The 
time since protostar formation is bracketed
by these two processes, yielding an age constraint of
$t \sim 8 \pm 4 \times 10^{4}$ yrs 
since formation of the YSO.
Due to the weak dependence of $T(r)$ on $L(t)$, this
timescale should be similar (within 0.5 - 1 dex) for a 
range of stellar masses. 

The combination of the warming envelope and material infall 
leads to small residence times of material in the warm gas, 
$\sim \mathrm{few} \times 10^{3} - 10^{4}$ yr, after which
the material is beam-diluted in a small structure such as a disk.
This naturally explains both the small variance in the ages
of hot cores inferred from chemical clocks, as well as the 
discrepant inferred ages between hot cores and their surrounding
halos.  Hot cores should disappear in $\sim 10^{5}$ yr,
once the YSO reaches the main sequence and ionizes the envelope.
These results suggest that massive star disks should contain most
of the complex species on scales of $<$ few hundred AU, which 
will be observable by the next generation of submm interferometers.

Finally, the
order of desorption kinetics assumed (zeroth vs. first-order) 
are not significant in comparison to source evolution or 
observational uncertainties.

\begin{acknowledgements}
      This work was partially supported under grants from 
      The Research Corporation (SDD).  Astrochemistry in Leiden
      is supported by the Netherlands Research School for
      Astronomy (NOVA) and by a Spinoza and a bezoekersbeurs 
      grant from the Netherlands
      Organization for Scientific Research (NWO).
\end{acknowledgements}

\end{document}